\documentstyle[aps,twocolumn,psfig]{revtex}
\begin{document}
\draft
\twocolumn[\hsize\textwidth\columnwidth\hsize\csname @twocolumnfalse\endcsname

\author {P.Pitanga}
\title{The Heavy Dirac Monopole}
\address{ Universidade Federal do Rio de Janeiro. Instituto de
F\'{\i}sica. Caixa Postal 68528\\
Cidade Universit\'aria. 21945-970. Rio de Janeiro, Brazil}
\date{december,2000}
\maketitle

\begin{abstract}  
We present a  model for the  Dirac magnetic monopole, sui\-ta\-ble for the strong coupling regime.
The  mag\-ne\-tic monopole is static, has charge $g$ and mass $M$, occupying a  vo\-lu\-me of radius $R\approx {\cal O}(g^2/M$). It is shown that inside each $n$-monopole there exist infinite multipoles. It is given an analytical proof of  the existence of monopole-antimonopole bound state. Theses bound-states might give additional  strong light to light scattering in the $p{\overline p}$ process  and in  $e^{+}e^{-}\rightarrow Z\rightarrow 3{\gamma}$ process.
\end{abstract}
\pacs{PACS number(s): 14.80.Hv, 12.90.+ b,13.85. Qk,13.85.Rm}
\vskip1pc]

In the present letter we extend the Dirac theory of pointlike  monopole,\cite {Di}, to the domain of high energy physics.
We have used the same global  Wu-Yang approach, adapted to the strong limit coupling,\cite {Wu}. The pa\-ra\-me\-tric equation of the vector potential ${\bf A}=(A_1,A_2,A_3)$, shows a singular self-energy with toric configurations inside. The corresponding potential $1$-form gives the magnetic field $2$-form as an infinite sum of Legendre Polynomials. The dependence of magnetic field ${\bf B}$, on the additional dimension ${\tau}$, gives the quantized magnetic flux  over a ${\pi}/2$-cycle. In the low energy limit, the theory simplifies
to the  Wu-Yang monopole theory.

The existence of the monopole is one of the open questions in  particle physics.
 If monopoles exist, then the elementary magnetic and electric charge $(g,e)$ are linked by the relation $ge=2{\pi}n $ where $n$ is an integer. If free quarks exist the magnetic charge is increased by a factor three.

A pair of high energy real photons may be produced from a virtual  monopole loop emerging from the  proton-antiproton collisions. The contribution of pointlike monopole to such diphoton production was calculated in \cite {Gi}.
 
 FermiLab researchers at the  D$0$ detector, have looked for signs of heavy  pointlike  monopole among the same data set used to discover the top quark \cite {Fe}.
 No evidence for the monopole was found  but lower limits on the mass of the monopole were established : $610$ GeV, $870$ GeV, or $1580$ GeV, if the pointlike Dirac monopole has spin $0$, $1/2$ or $1$.

A similar monopole loop production occurs also in the process $e^{+}e^{-} \rightarrow Z\rightarrow 3{\gamma}$ and was tested at the CERN $e^{+}e^{-}$ collider LEP \cite {Le}.
In  Fermilab's paper it was stressed that further theoretical work is desirable, to upgrade the theory of pointlike monopoles. One of the reasons is that the non-observation of a {\it new extra dimension} in QED implies that the monopole mass should exceed $100$ TeV. Existing or planned particle accelerators will not have enough energy to produce such monopole.

Recently new features on the observable effects of the virtual monopole loop were reported in \cite {Gii}, considering only pointlike monopole. In that paper there is a call for a new theory of heavy monopole  having  the standard $SU(2)\times U(1)$ theory as the lower energy limit. This
challenging statement is one of the main motivations for our work. 

We begin by recalling the fact that the strong coupling regime is incompatible with pointlike structure of the particles, so the monopole must have a definite volume  of radius $ R\approx{\cal O}( g^2/M)$ to accommodate the self energy of the strong regime coupling. The size of the monopole being large, compared with the quantum length scale, permits a classical description. 

First of all we have to answer the following question:{\it If the monopole  occupies a volume of radius $R\approx {\cal O}(g^2/M)$, what class of  compact $3$-D space ${\Omega}$,  
 is capable to enclose the monopole such that $ \oint_{\partial {\Omega}} {\bf B}=4{\pi}g$,? } 

The following  statements were  keys  to discover the appropriate ${\Omega}$: 
\begin{enumerate}
\item  The magnetic charge density cannot be a smooth distribution. If it were, then the Maxwell equations would be substituted by the Yang-Mills-Higgs equations, to allow finite-energy action monopole solutions. \cite {Ta}.
\item The quantum consistency for QED or $SU(2)\times U(1)$, requires that  the magnetic charge cannot be  spread out over a length scale larger than the minimum for which the standard model is accurate \cite {As}.  
\item The magnetic charges must be  compacted  to produce the ``cones over cones'' stratification of the  configuration space of the Yang-Mills-Higgs $SU(2)\times U(1)$ theory on ${\bf R}^3\times S^1$ space,\cite {Jf,Ma}. 
\end{enumerate}

 We consider ${\Omega}$ a regular spatial domain bounded by a surface ${\partial {\Omega}}$.  By ``regular'' we mean that ${\Omega}$ may have an infinite number of isolated singularities. 

The exterior space is the complement in the Euclidean space ${\cal E}={\bf R}^3-{\Omega}$. The spaces ${\cal E}$ and ${\Omega}$ have  a common boundary, homeomorphic to a Riemannian sphere $S^2$. The region ${\Omega}$, in contrast to ${\cal E}$, must be as small as possible  to describe the strong coupling regime. We have found that in the strong coupling regime, the geometry of ${\Omega}$ must be sub-Riemannian  to satisfy the  above  requirements. Euclidean geometry holds locally in  ${\cal E}$, where the monopole can be considered pointlike, in the low energy limit.
 
We have chosen the geometry of the compact $3$-D Heisenberg group ${\cal H}(3)$  as the geometry of ${\Omega}$. We have done so for several reasons. The non-compact $3$-D Heisenberg group $H(3)$,(also known as Weyl group \cite {Rg}) is the  only connected nilpotent non-Abelian Lie group, homeomorphic (as manifold) to ${\bf R}^3$, \cite {Ge}. It is represented by upper triangular matrices $a=(x_1,x_2,x_3)$ with one in  each  diagonal entry. The group product is defined as $a \cdot a'=(x_1+x'_1,x_1+x'_2,x_3+x'_3+x'_2x_1-x'_1x_2)$. The discrete subgroup $D$, is generated by the  matrices $D_n= (1,0,0),(0,1,0),(0,0,1/m)$, where $m=1,2,...,$. The compact group is the coset ${\cal H}(3)=H(3)/D$. As a compact Lie group, ${\cal H}(3)$, has an invariant Haar measure  which provides a small finite volume, \cite {Ss}. Its Lie algebra satisfies the relations
\begin{equation}
[X_1,X_2]=X_3;\;\;[X_3,X_1]=0;\;[X_3,X_2]=0
\end{equation} Therefore, due the nointegrability, there exist a  vector field in this Lie algebra with undefined line integral, $\oint{\bf A}\cdot d{\bf x}\neq 0$, around an unshrinkable loop
\cite {Pit}.

 The $3$-D Heisenberg  ball has a complicated  boundary \cite {Aa,El}, ${\partial {\Omega}}={\partial {\Omega}}_{0} \bigcup {\partial {\Omega}}^{+}_{m}{\bigcup}{\partial {\Omega}}^{-}_{m}$. The boundary extends to the interior with an infinite number of smooth small cones with two covers, ${\partial {\Omega}}^{\pm}_{ m}$, $m=1,2,...$. These cones are the charges of the $SU(2)$ theory. In our model, each small cone may receive  a  magnetic charge $g=\pm 2{\pi}n/e$ along its surface. The great exterior boundary ${\partial {\Omega}}_0 $, extends to the interior and terminates at two conic singularities  ( the north and south pole). This boundary is topologically equivalent to a double  covering of a Riemannian sphere $S^2$.

In the Heisenberg ball the distance  scales as
$d_{{\cal H}}(S(X_1),S(X_2))={\lambda}d_{{\cal H}}(X_1,X_2)$, the volume as
${\lambda}^4$,\cite {Ab}. The relation between the Heisenberg distance and the Euclidean distance is $d_{{\cal H}}=\sqrt {d_E}$. Therefore the maximal radius of the monopole is ${\varepsilon}=g/M^{1/2}$. 

The shape of the boundary ${\partial {\Omega}}$, is determined  by the set of end points of the geodesics of ${\Omega}$. 
The geodesics of the Heisenberg ball are given by the minimizers of the Heisenberg distance in the interval $[0,{\tau}]$:

\begin{mathletters}
\label{e}
\begin{equation}
\label{21}
d_{{\cal H}(3)}=\int_0^{\tau} \sqrt{({\dot x}_1^2+{\dot x}^2_2 +{\dot x}^2_3)}dt 
\end{equation}
\begin{equation}
\label{rr}
{\dot x}_3+x_1{\dot x}_2-x_2 {\dot x}_1=0
\end{equation}
\end{mathletters}

The set of end points of the geodesics of the  system (\ref {e}), is given
by the parametric equations,\cite {e}: 
\begin{mathletters}
\label{111}
\begin{equation}
\label{303}
x_1={\varepsilon}\left[\frac{sin({\theta} +{\phi})-sin{\phi}}{{\theta}}\right]
\end{equation}
\begin{equation}
\label{444}
 x_2={\varepsilon}\left[\frac{-cos{\phi} +cos({\theta} +{\phi})}{{\theta}}\right]
\end{equation}
\begin{equation}
\label{555}
x_3= {\varepsilon}^2\left[\frac{{\theta} -sin {{\theta}}}{{\theta}^2}\right]
\end{equation}
\end{mathletters}
 $$- 2(k+1){\pi}\leq{\theta}\leq 2(k+1){\pi};\;\;k=0,1,2,..;\;\;\;0\leq{\phi}\leq 
2{\pi}$$
where ${\theta}\equiv{\theta}_{\cal H}=2{\pi}{\tau}$, ${\tau}\in {\bf R}/[0,1]\equiv S^1$.

In fig.1 is shown parametric plot of the north-hemisphere ${\partial {\Omega}}^{+}$, with three magnetic charges $0\leq{\theta}\leq 8{\pi}$. In fig.2 is shown in detail one magnetic charge, $2{\pi}\leq {\theta}\leq 4{\pi}$.

Let us now consider the map :
${\bf A}:{\partial{\Omega}}^{+}\rightarrow {\partial {\cal S}}\subset {\cal M}$, where ${\cal M}$ is the space of all potential vectors modulo gauge transformation. In order to work with the three components of ${\bf A}=(A_1,A_2,A_3)$ we impose the gauge constraint ${\nabla}\cdot{\bf A}=0$, and $V=0$ for the scalar potential.  

The energetic configuration space ${\cal S}$, is compact and has also  
  a complicated  boundary ${\partial {\cal S}}={\partial {\cal S}}_0\bigcup T_{m}$ . 
The boundary  ${\partial {\cal S}}_0$ has  a pear-like shape without north-pole, with a singularity at the south-pole. This configuration is the self-energy of the monopole. The other configurations  $T_{m}$, inside ${\partial {\cal S}}_0$, are ``tori inside tori'' in harmonic correspondence with the ``cones over cones'' structures inside ${\partial {\Omega}}^{+}_0$.

 As in the Hopf map between Riemannian spheres, we identify the great circles in ${\partial {\Omega}}^{+}_0$, with points in ${\partial {\cal S}}_0$, \cite {St}. In the present case we identify the equator of ${\partial {\Omega}}^{+}_0$ (${\theta}=0$) with the  polar circle of ${\partial {\cal S}}_0$. The  singularity at the north of ${\partial {\Omega}}^{+}_0$ ( ${\theta}=2{\pi}$), with the singularity at the south of ${\partial {\cal S}}_0$. 
In order to obtain such correspondence the components of the potential vector ${\bf A}$ must satisfy the relation
\begin{equation}
\label{to}
A_1^2+A_2^2+A_3^2= M(x_1^2(2{\theta}) +x_2^2(2{\theta}))
\end{equation}
where we have taken into account that one turn in the meridional plane of ${\partial {\Omega}}^{+}_0$, corresponds to two turns in the meridional plane of ${\partial {\cal S}}_0$. Using equations (\ref {303}) and (\ref {444}) in (\ref {to}),
 we obtain the parametric equations of the potentials for $r_{0}<r\leq  {\varepsilon}$:
\begin{mathletters}
\label{333}
\begin{equation}
\label{331}
A_1=\frac {g}{\sqrt 2}\frac{\sqrt{1-cos2{\theta}}}{{\theta}}\sin{\theta}cos{\phi}\end{equation}
\begin{equation}
\label{332}
A_2=\frac{g}{\sqrt 2}\frac{\sqrt{1-cos2{\theta}}}{{\theta}}\sin{\theta}sin{\phi}
\end{equation}
\begin{equation}
\label{330}
A_3=\frac{g}{\sqrt 2}\frac{\sqrt{1-cos2{\theta}}}{{\theta}}\\\\cos{\theta}
\end{equation}
\end{mathletters}
$$-(k+1){\pi}\leq{\theta}\leq  (k+1){\pi};\;\;\;0\leq {\phi}\leq 2{\pi};\;\;k=0,1, 2,...$$
Equations (\ref {333}) are the parametric equations of the energetic configuration corresponding to the charge distribution on the Heisenberg ball ${\Omega}$. The positive branch of these solutions is
associated with the monopole, while the negative branch is associated with the anti-monopole, like in the electron-positron theory \cite {G,Pi}.  In fig. 3 is shown the  parametric plot of the self-energy, $0\leq {\theta}\leq {\pi}$. In fig.4 is show the configuration corresponding to three magnetic charges, $0\leq{\theta}\leq 4{\pi}$. 
The two branched solutions are connected through the branch cut in the complex plane beginning at the threshold for pair productions. Bellow and near this point, there exist bound states of the scattering matrix connecting $-(1+k){\pi}$ solutions to $(1+k){\pi}$ solutions. In fig.5 it is shown the parametric plot of a bound-pair between $-{\pi}\leq{\theta}\leq 2{\pi}$.

To obtain the magnetic field of such configurations  we must write the potentials $1$-forms, for $r_0<r\leq {\varepsilon}$, where $r_0$ is the length beyond of which we must use the Bogomolny equations of the $SU(2)\times U(1)$ Yang-Mills-Higgs theory. 

The Jacobian of (\ref {333}) gives the Haar measure of  ${\cal S}$:
${\omega}({\theta})=2{\pi}r^2(sin^4{\theta}/{\theta}^3)d{\theta}\equiv 2{\pi}r^2f({\theta})d{\theta}$. Let us take $A_r=A_{\theta}=0$, and
\begin{equation}
\label{330}
A_{\phi}=\frac{g}{\sqrt 2}\frac{1}{rf({\theta})}
\sqrt{1-cos2{\theta}}
\end{equation}
 It is easy to verify that $\lim_{{\theta}\rightarrow 0}A_{\phi}=g/r$ and $\lim_{{\theta}\rightarrow  {\pi}}A_{\phi}={\infty}$. With the above choice, the phase factor of the gauge theory becomes well defined  around the singularity. Indeed, the integral of the vector potential around the loop  defined by the line element $d{\bf x}=2{\pi}rf({\theta})d{\theta}{\hat{\phi}}$, gives 
\begin{equation}
\label{330}
\oint {\bf A}\cdot d{\bf x}=2{\pi}\frac{g}{\sqrt 2}\int_{0}^{{\pi}}
\sqrt {1-cos2{\theta}}d{\theta}=4{\pi}g
\end{equation}
  
Thus the potential $1$-form  must be
\begin{equation}
\label{330}
{\bf A}^{+}=\frac {2g}{{\sqrt {2}}r}(\sqrt{1-cos2{\theta}})d{\phi};\;\;0\leq {\theta}<(1+k){\pi}
\end{equation}
corresponding to monopole solutions. 

The magnetic field  of the monopole is given by
\begin{equation}
{\bf B}^{+}=d{\bf A}^{+}=\frac {4g}{{\sqrt {2}}}\left(\frac{\sin{\theta}\\\\cos{\theta}}{\sqrt{1-cos2{\theta}}}\right)d{\theta}\wedge d{\phi}
\end{equation}
or
\begin{equation}
\label{33}
{\bf B}^{+}=4g\left(\\sin{\theta}\\\\cos{\theta}\right)
\left(\sum_{k=0}^{\infty}P_k(\cos2{\theta})\right)d{\theta}\wedge d{\phi}
\end{equation}
where $P_k(cos2{\theta})$ is the Legendre polynomial. In fig.6 is shown the parametric plot of the magnetic field in  the ''instant'' ${\tau}=1$,  considering the first two terms of the expansion $k=0,k=1$.

We remark that the modulator factor is  the field of the Wu-Yang monopole, the field of the self-energy.

The integration of (\ref {33}), over a ${\pi}/2$-cycle, gives the  flux  associated to the magnetic field. 
 It is easy to verify, that only the self-energy is responsible to the  flux. Using the orthogonality properties of the Legendre Polynomials, we obtain
\begin{equation}
\int_{0}^{{\pi}/2}\left(\sin{\theta}\cos{\theta}\right)
\left(\sum_{k=0}^{\infty}P_k(cos2{\theta}({\tau}))\right)d{\theta}=\frac{1}{2}
\end{equation}
 So the flux  of the $n$-monopole is:
\begin{equation}
{\Phi}^{+}=\oint_{{\partial {\Omega}}} {\bf B}^{+}= 4{\pi}g= \frac {8{\pi}^2n}{e};\;\;n=1,2,...
\end{equation}
In conclusion, we have shown that the extension of the pointlike monopole to strong coupling regime is the ${\cal H}(3)\times SU(2)\times U(1)$ - gauge theory, where ${\cal H}(3)$ is the compact $3$-D Heisenberg group. One of the features of our  theory, not present in the pointlike theory, is the bound state solutions with  toroidal distribution of energy around a very thin  tube of quantized flux, as  shown in fig.5. The bound pair 
is confined. Indeed, due to the flux tube linking the poles, any tentative to separate them leads to the energy rise, because by construction this configuration is the minimum energy.
Thus if the virtual monopole loop exists, then a Bohm-Aharonov effect might   change the momentum  of the protons (antiprotons) in the transverse direction by just a quantum of magnetic flux, like in the Bohm-Aharonov experiment for electrons \cite {Rc}.

Finally we remark that the Haar measure of the ${\cal H}(3)$ group reveals,  structures inside structures, measurable depending on the energetic scale used to measure,\cite {We1}.

\begin{figure} 
\psfig{file=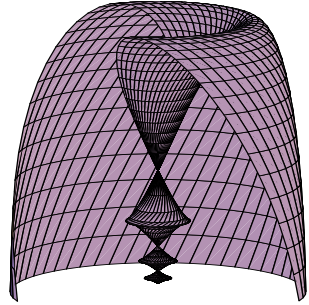,height=4.cm,width= 4.cm}
\caption{${\partial {\Omega}}^{+}$ with three charges.$-0\leq{\theta}\leq 8{\pi};\;\;0\leq{\phi}\leq 1.3{\pi}$}
\end{figure}
\begin{figure}
\psfig{file=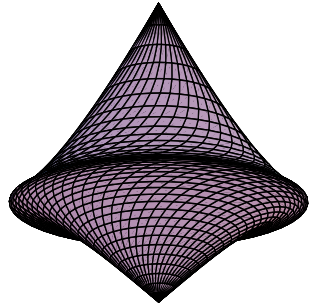,height=4.cm,width= 4.cm}
\caption{The magnetic charge.$2{\pi}\leq{\theta}\leq 4{\pi};\;\;0\leq{\phi}\leq 2{\pi}$}
\end{figure}
\begin{figure}
\psfig{file=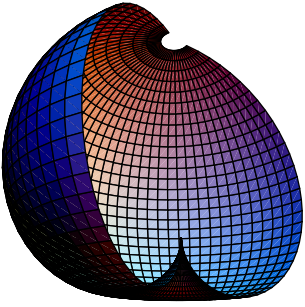,height=4.cm,width= 4.cm}
\caption{The self-energy.$0\leq{\theta}\leq {\pi};\;\;0\leq{\phi}\leq 1.3{\pi}$}
\end{figure}
\begin{figure} 
\psfig{file=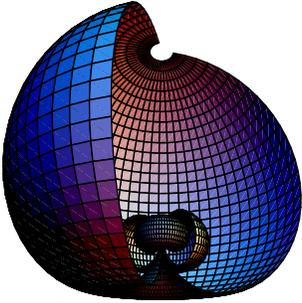,height=4.cm,width= 4.cm}
\caption{Energetic configurations of three-charges. $0\leq{\theta}\leq 4{\pi};\;\;0\leq{\phi}\leq 1.3{\pi}$}
\end{figure}
\begin{figure} 
\psfig{file=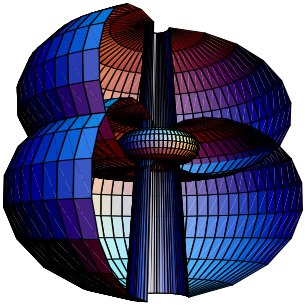,height=4.cm,width= 4.cm}
\caption{ A Bound-states. $-{\pi}\leq{\theta}\leq 2{\pi};\;\;0\leq{\phi}\leq 1.3{\pi}$}
\end{figure}
\begin{figure} 
\psfig{file=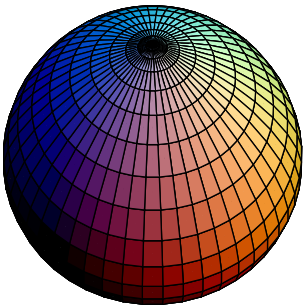,height=4.cm,width= 4.cm}
\caption{The monopole field in the ``instant''${\tau}=1$}
\end{figure}

\vspace{1cm}
This paper is dedicated to the memory of Professor Guido Beck and Professor Carlos Marcio do Amaral.

\end{document}